\documentclass[12pt,preprint]{aastex}

\def\aj{AJ}%
\def\apj{ApJ}%
\def\apjs{ApJS}%
\def\ao{Appl.~Opt.}%
\def\aap{A\&A}%
\def\mnras{MNRAS}%
\def\pasp{PASP}%
\def\nat{Nature}%

\slugcomment{Revised version}

\shorttitle{Observations of $\eta$ Crv with the Large Binocular Telescope Interferometer}
\shortauthors{Defr\`ere et al.}

\begin{document}

\title{First-light LBT nulling interferometric observations: warm exozodiacal dust resolved within a few AU of $\eta$\,Crv}

\author{D.~Defr\`ere\altaffilmark{1}, P.M.~Hinz\altaffilmark{1}, A.J.~Skemer\altaffilmark{1}, G.M.~Kennedy\altaffilmark{2}, V.P.~Bailey\altaffilmark{1}, W.F.~Hoffmann\altaffilmark{1},  B.~Mennesson\altaffilmark{3}, R.~Millan-Gabet\altaffilmark{4}, W.C.~Danchi\altaffilmark{5},  O.~Absil\altaffilmark{6,}\footnote{F.R.S.-FNRS Research Associate.} ,  P.~Arbo\altaffilmark{1}, C~ Beichman\altaffilmark{4}, G.~Brusa\altaffilmark{1}, G.~Bryden\altaffilmark{3},  E.C.~Downey\altaffilmark{1}, O.~Durney\altaffilmark{1}, S.~Esposito\altaffilmark{7}, A.~Gaspar\altaffilmark{1}, P.~Grenz\altaffilmark{1}, C.~Haniff\altaffilmark{8}, J.M.~Hill\altaffilmark{9}, J.~Lebreton\altaffilmark{4}, J.M.~Leisenring\altaffilmark{1}, J.R.~Males\altaffilmark{1,}\footnote{NASA Sagan Fellow.} , L.~Marion\altaffilmark{6}, T.J.~McMahon\altaffilmark{1}, M.~Montoya\altaffilmark{1}, K.M.~Morzinski\altaffilmark{1,\dagger}, E.~Pinna\altaffilmark{7}, A.~Puglisi\altaffilmark{7}, G.~Rieke\altaffilmark{1}, A.~Roberge\altaffilmark{5}, E.~Serabyn\altaffilmark{3}, R.~Sosa\altaffilmark{1}, K.~Stapeldfeldt\altaffilmark{5}, K.~Su\altaffilmark{1}, V.~Vaitheeswaran\altaffilmark{1}, A.~Vaz\altaffilmark{1}, A.J.~Weinberger\altaffilmark{10}, and M.C.~Wyatt\altaffilmark{2}}

\affil{\altaffilmark{1}Steward Observatory, Department of Astronomy, University of Arizona, 933 N. Cherry Ave, Tucson, AZ 85721, USA}
\affil{\altaffilmark{2}Institute of Astronomy, University of Cambridge, Madingley Road, Cambridge CB3 0HA, UK}
\affil{\altaffilmark{3}Jet Propulsion Laboratory, California Institute of Technology, 4800 Oak Grove Drive, Pasadena CA 91109-8099, USA}
\affil{\altaffilmark{4}NASA Exoplanet Science Institute, California Institute of Technology, 770 South Wilson Avenue, Pasadena CA 91125, USA}
\affil{\altaffilmark{5}NASA Goddard Space Flight Center, Exoplanets \& Stellar Astrophysics Laboratory, Code 667, Greenbelt, MD 20771, USA}
\affil{\altaffilmark{6}D\'epartement d'Astrophysique, G\'eophysique et Oc\'eanographie, Universit\'e de Li\`ege, 17 All\'ee du Six Ao\^ut, B-4000 Sart Tilman, Belgium}
\affil{\altaffilmark{7}INAF-Osservatorio Astrofisico di Arcetri, Largo E. Fermi 5, I-50125 Firenze, Italy}
\affil{\altaffilmark{8}Cavendish Laboratory, University of Cambridge, JJ Thomson Avenue, Cambridge CB3 0HE, UK}
\affil{\altaffilmark{9}Large Binocular Telescope Observatory, University of Arizona, 933 N. Cherry Ave, Tucson, AZ 85721, USA}
\affil{\altaffilmark{10}Department of Terrestrial Magnetism, Carnegie Institution of Washington, 5241 Broad Branch Road NW, Washington, DC, 20015, USA}

\email{ddefrere@email.arizona.edu}

\begin{abstract}
We report on the first nulling interferometric observations with the Large Binocular Telescope Interferometer (LBTI), resolving the N' band (9.81 - 12.41\,$\mu$m) emission around the nearby main-sequence star $\eta$\,Crv  (F2V, 1-2\,Gyr). The measured source null depth amounts to 4.40\% $\pm$ 0.35\% over a field-of-view of 140\,mas in radius ($\sim$2.6\,AU at the distance of $\eta$\,Crv) and shows no significant variation over 35$^\circ$ of sky rotation. This relatively low null is unexpected given the total disk to star flux ratio measured by \emph{Spitzer/IRS} ($\sim$23\% across the N' band), suggesting that a significant fraction of the dust lies within the central nulled response of the LBTI (79\,mas or 1.4\,AU). Modeling of the warm disk shows that it cannot resemble a scaled version of the Solar zodiacal cloud, unless it is almost perpendicular to the outer disk imaged by \emph{Herschel}. It is more likely that the inner and outer disks are coplanar and the warm dust is located at a distance of 0.5-1.0\,AU, significantly closer than previously predicted by models of the IRS spectrum ($\sim$3\,AU). The predicted disk sizes can be reconciled if the warm disk is not centrosymmetric, or if the dust particles are dominated by very small grains. Both possibilities hint that a recent collision has produced much of the dust. Finally, we discuss the implications for the presence of dust at the distance where the insolation is the same as Earth's (2.3\,AU).
\end{abstract}
\section{Introduction}
\label{sec:intro}

The possible presence of dust in the habitable zones of nearby main-sequence stars is considered a major threat for the direct imaging and characterization of Earth-like extrasolar planets (exo-Earths) with future dedicated space-based telescopes. Several independent studies have addressed this issue and concluded that visible to mid-infrared direct detection of exo-Earths would be seriously hampered in the presence of dust disks 10 to 20 times brighter than the Solar zodiacal cloud assuming a smooth brightness distribution \citep[e.g.,][]{Beichman:2006b,Defrere:2010,Roberge:2012}. The prevalence of exozodiacal dust at such a level in the terrestrial planet region of nearby planetary systems is currently poorly constrained and must be determined to design these future space-based instruments. So far, only the bright end of the exozodi luminosity function has been measured on a statistically meaningful sample of stars \citep[][]{Lawler:2009,Kennedy:2013}. Based on \emph{WISE} observations and extrapolating over many orders of magnitude, \citet[][]{Kennedy:2013} suggest that at least 10\% of Gyr-old main-sequence stars may have sufficient exozodiacal dust to cause problems for future exo-Earth imaging missions. To determine the prevalence of exozodiacal dust at the faint end of the luminosity function, NASA has funded the Keck Interferometer Nuller (KIN) and the Large Binocular Telescope Interferometer (LBTI) to carry out surveys of nearby main sequence stars. Science observations with the KIN started in 2008 and the results  were reported recently \citep[][]{Millan-gabet:2011,Mennesson:2014}. One of their analyses focused on a sample of 20 solar-type stars with no far infrared excess previously detected (i.e, no outer dust reservoir). Assuming a log-normal luminosity distribution,  they derived the median level of exozodiacal dust around such stars to be below 60 times the solar value with high confidence (95\%). Yet, the state-of-the-art exozodi sensitivity achieved per object by the KIN is approximately one order of magnitude larger than that required to prepare future exo-Earth imaging instruments.

The LBTI is the next step. A survey, called the Hunt for Observable Signatures of Terrestrial Planetary Systems \cite[HOSTS,][]{Hinz:2013,Danchi:2014}, will be carried out on 50 to 60 carefully chosen nearby main-sequence stars over the next 4 years (Weinberger et al. in prep). In this paper, we report on the first mid-infrared nulling interferometric observations with the LBTI obtained on 2014 February 12 as part of LBTI's commissioning. We observed the nearby main-sequence star $\eta$\,Crv (F2V, 1.4\,M$_\odot$, 18.3\,pc), previously known to harbor unusually high levels of circumstellar dust. The infrared excess was first detected with \emph{IRAS} \citep[][]{Stencel:1991} and confirmed later by \emph{Spitzer/MIPS} at 70\,$\mu$m \citep[][]{Beichman:2006}, \emph{Spitzer/IRS} at 5 -- 35\,$\mu$m \citep[][]{Chen:2006,Lisse:2012}, and \emph{Herschel} at 70 -- 500\,$\mu$m \citep{Duchene:2014}. Interestingly, the SED exhibits two distinct peaks which can be explained by the presence of two well separated dust belts. The cold outer one was first resolved in the submillimeter and found to lie at a distance of $\sim$150\,AU from the star \cite[][]{Wyatt:2005}. Mid-infrared interferometric observations with VLTI/MIDI and the KIN resolved the inner belt to be located within 3\,AU of the host star \cite[][]{Smith:2008,Smith:2009,Millan-gabet:2011}, a conclusion supported by blackbody models that place the warm dust near 1\,AU \citep[e.g.,][]{Wyatt:2005}. A detailed analysis of the dust composition in the inner belt using the \emph{Spitzer/IRS} spectrum suggests a location nearer to 3\,AU due to the presence of grains as small as 1\,$\mu$m that are warmer than blackbodies for the same stellocentric distance \citep[][]{Lisse:2012}. Most importantly, the mid-infrared excess emission is remarkably large and intriguing considering the old age of the system \citep[1-2\,Gyr,][]{Ibukiyama:2002,Mallik:2003,Vican:2012} and all the aforementioned studies agree that it cannot be explained by mere transport of grains from the outer belt \citep[at least not without a contrived set of planetary system parameters,][]{Bonsor:2012}. This rather supports a rare and more violent recent event \citep[e.g.,][]{Gaspar:2013} and makes $\eta$\,Crv an ideal object to study catastrophic events like the Late Heavy Bombardment that might have happened in the early solar system \citep{Gomes:2005} or a recent collision between planetesimals \cite[e.g., as postulated for BD\,+20307,][]{Song:2005}.

\section{Observations and data reduction}\label{sec:instr}
\subsection{Instrumental setup}

The Large Binocular Telescope consists of a two 8.4-m aperture optical telescopes on a single ALT-AZ mount installed on Mount Graham in southeastern Arizona (at an elevation of 3192 meters) and operated by an international collaboration among institutions in the United States, Italy, and Germany \citep{Hill:2014,Veillet:2014}. Both telescopes are equipped with deformable secondary mirrors which are driven with the LBT's adaptive optics system to correct atmospheric turbulence at 1 kHz \citep{Esposito:2010, Bailey:2014}. Each deformable mirror uses 672 actuators that routinely correct 400 modes and provide Strehl ratios exceeding 80\%, 95\%, and 99\% at 1.6\,$\mu$m, 3.8\,$\mu$m, and 10\,$\mu$m, respectively \citep{Esposito:2012,Skemer:2014}. The LBTI is an interferometric instrument designed to coherently combine the beams from the two 8.4-m primary mirrors of the LBT for high-angular resolution imaging at infrared wavelengths \cite[1.5-13\,$\mu$m,][]{Hinz:2012}. It is developed and operated by the University of Arizona and based on the heritage of the Bracewell Infrared Nulling Cryostat on the MMT \cite[BLINC,][]{Hinz:2000}. The overall LBTI system architecture and performance will be presented in full detail in a forthcoming publication (Hinz et al.\ in prep). In brief,  the LBTI consists of a universal beam combiner (UBC) located at the bent center Gregorian focal station and a cryogenic Nulling Infrared Camera (NIC). The UBC provides a combined focal plane for the two LBT apertures while the precise overlapping of the beams is done in the NIC cryostat. Nulling interferometry, a technique proposed 36 years ago to image extra-solar planets \citep{Bracewell:1978}, is used to suppress the stellar light and improve the dynamic range of the observations. The basic principle is to combine the beams in phase opposition in order to strongly reduce the on-axis stellar light while transmitting the flux of off-axis sources located at angular spacings which are odd multiples of 0.5$\lambda/B$ (where $B=14.4$\,m is the distance between the telescope centers and $\lambda$ is the wavelength of observation). Beam combination is done in the pupil plane on a 50/50 beamsplitter which can be translated to equalize the pathlengths between the two sides of the interferometer. One output of the interferometer is reflected on a short-pass dichroic and focused on the NOMIC camera \citep[Nulling Optimized Mid-Infrared Camera,][]{Hoffmann:2014}. NOMIC uses a 1024x1024 Raytheon Aquarius detector split into 2 columns of 8 contiguous channels. The optics provides a field of view (FOV) of 12~arcsecs with a plate-scale of  0.018~arcsecs. Tip/tilt and phase variations between the LBT apertures are measured using a fast-readout (1\,Hz) K-band PICNIC detector (PHASECam) which receives the near-infrared light from both outputs of the interferometer. Closed-loop correction uses a fast pathlength corrector installed in the UBC (see more details in Defr\`ere et al.\ 2014). 

\subsection{Observations}

Nulling interferometric observations of $\eta$\,Crv were obtained in the N' band (9.81 - 12.41\,$\mu$m) on 2014 February 12 as part of LBTI's commissioning. The basic observing block (OB) consisted of one thousand frames, each having an integration time of 85 ms, for a total acquisition time of 110 seconds including camera overheads. The observing sequence was composed of several successive OBs at null, i.e.\ with the beams from both telescopes coherently overlapped in phase opposition, and one OB of photometric measurements with the beams separated on the detector. In order to estimate and subtract the mid-IR background, the OBs at null were acquired in two telescope nod positions separated by 2\farcs{3} on the detector. We acquired 3 different observations of $\eta$\,Crv interleaved with 4 observations of reference stars to measure and calibrate the instrumental null floor (CAL1-SCI-CAL2-SCI-CAL1-SCI-CAL3 sequence; see target and calibrator information in Table~\ref{tab:calib}). To minimize systematic errors, calibrator targets were chosen close to the science target, both in terms of sky position and magnitude, using the \emph{SearchCal} software \citep{Bonneau:2011}. The 7 observations  occurred over a period of approximately 3 hours around the meridian transit (see \textit{u-v} plane covered by the science observations in Fig.~\ref{fig:TF}) with relatively stable weather conditions. The seeing was $\sim$1\farcs{4} for the first observation and 1\farcs{0}-1\farcs{2} for the remainder observations. The adaptive optics systems (AO) were locked with 300 modes (left side) and 400 modes (right side) at a frequency of 990\,Hz over the duration of the observations\footnote{The different number of modes used on each telescope has a negligible impact on the null depth (i.e., a few $10^{-7}$). Furthermore, this effect is constant during the night and calibrates out completely.}. Fringe tracking was carried out at 1kHz using a K-band image of pupil fringes and an approach equivalent to group delay tracking \citep[][]{Defrere:2014}. The phase setpoint was optimized at least once per observation to take out any optical path variation between the K-band, where the phase is measured and tracked, and the N'-band, where the null depth is measured (making sure that the phase maintained at K-band provided the best possible nulls at N' band). The NOMIC detector was used with the small well depth for better sensitivity and in subframe mode (512$\times$512 pixels) to reduce the camera overhead.

%Such a variation generally comes from varying longitudinal dispersion between the two interfering light beams and determines the position of the best null as a function a wavelength. In the case of the LBTI,  the imbalance is dominated by the differential water vapor column above the 2 telescopes. 

\begin{table*}[!t]
\begin{center}
\caption{Basic properties of $\eta$\,Crv and its calibrators.}\label{tab:calib}
    \begin{tabular}{c c c c c c c c c c}
        \hline
        \hline
        ID & HD & RA-J2000 & DEC-J2000 & Type & $m_V$ & $m_K$  & $F_{\rm \nu,N'}$ & $\theta_{\rm LD}\pm1\sigma$ & Refs. \\
        & & [d m s] & [d m s] & & & & [Jy] & [mas] & \\
        \hline
        $\eta$\,Crv & 109085 & 12 32 04 & -16 11 46 & F2V  & 4.30 & 3.37 & 1.76 & 0.819 $\pm$ 0.119 & A13\\
         CAL 1  & 108522 & 12 28 02 & -14 27 03 & K4III & 6.80 & 3.46 & 1.68 & 1.204 $\pm$ 0.016 & M05\\
         CAL 2  & 107418 & 12 20 56 & -13 33 57 & K0III & 5.15 & 2.83 & 2.96 & 1.335 $\pm$ 0.092 & B11\\
         CAL 3  & 109272 & 12 33 34 & -12 49 49 & G8III & 5.59 & 3.60 & 1.42 & 0.907 $\pm$ 0.063 & B11\\
    \hline
    \end{tabular}\\
\end{center}
    {\small References. Coordinates, spectral types and V/K magnitudes from SIMBAD; N'-band flux densities computed by SED fit (Weinberger et al.\ in prep); Limb-darkened diameters and 1-$\sigma$ uncertainties from [A13] \cite{Absil:2013}, [M05] \cite{Merand:2005}, and [B11] \cite{Bonneau:2011}.}
\end{table*}

\subsection{Data reduction and calibration}

\begin{figure}[!t]
\centering
\includegraphics[height=7.0cm]{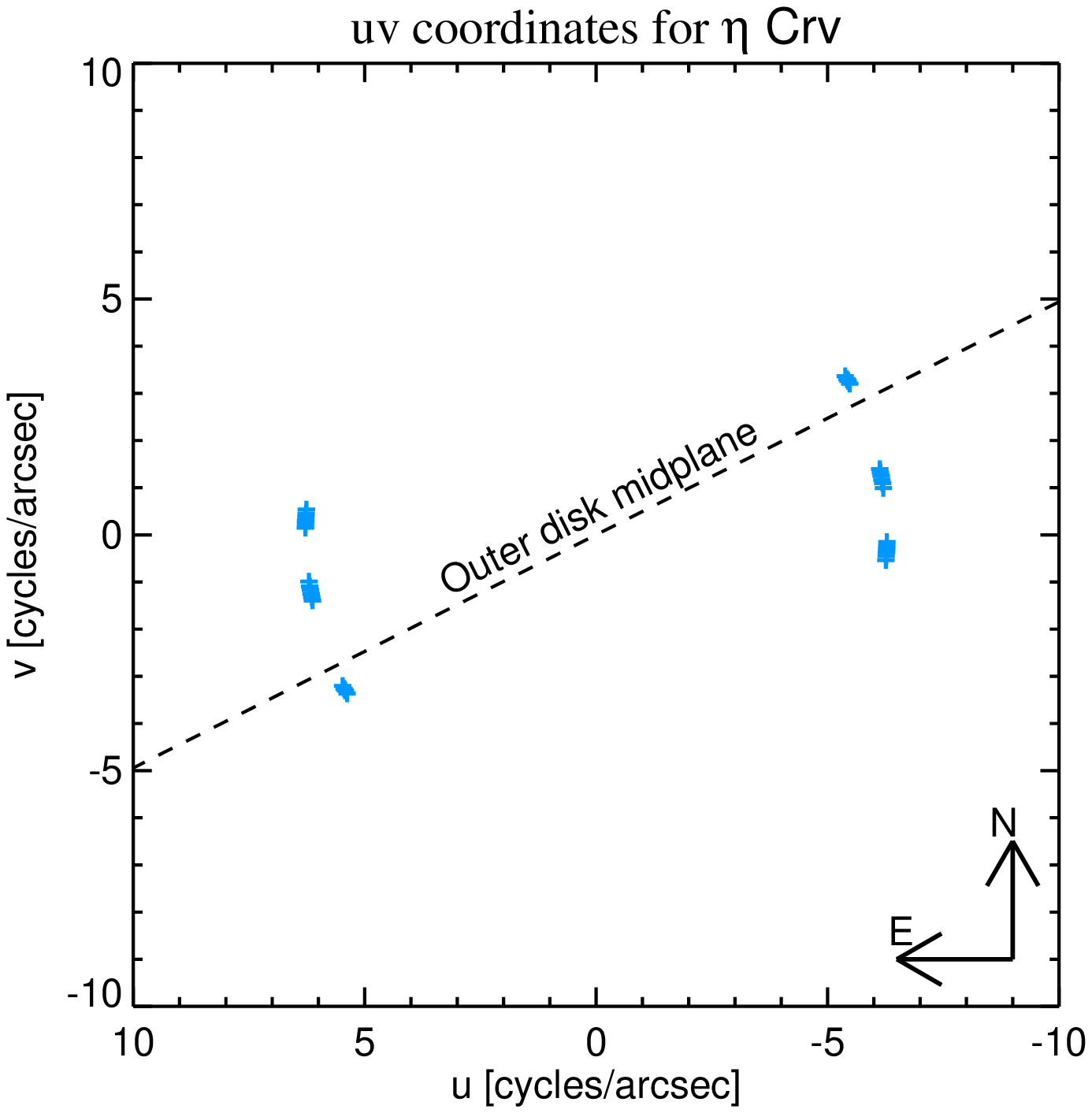}
\includegraphics[height=7.0cm]{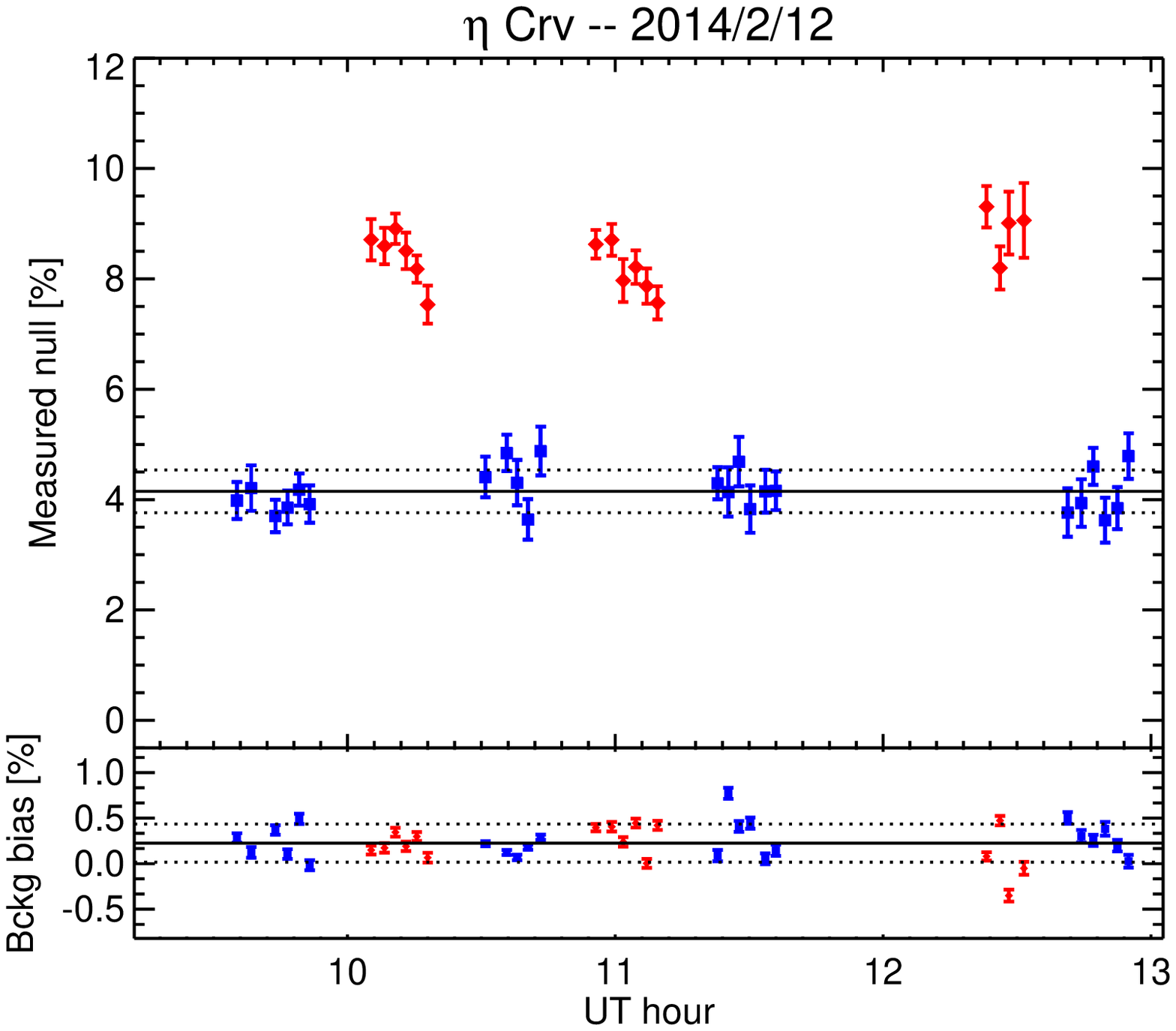}
\caption{Left, sampling of the Fourier (\textit{u,v}) plane obtained for $\eta$\,Crv on 2014 February 12. Each blue point and its centrosymmetric counterpart represent the orientation of the LBTI baseline for a given OB. The orientation of the outer disk midplane \citep[i.e., 116\fdg3,][]{Duchene:2014} is represented by the black dashed line. The right figure shows the corresponding raw null measurements per OB as a function of UT time (top panel). The blue squares show the calibrator measurements while the red diamonds represent the $\eta$\,Crv measurements. The estimated instrumental null floor is represented by the solid black line and the corresponding 1-$\sigma$ uncertainty by the dotted lines. The bottom panel shows the corresponding background error estimate measured in a nearby empty region of the detector. }
\label{fig:TF}
\end{figure}

Data reduction and calibration were performed using the \emph{nodrs} pipeline developed by the LBTI and HOSTS teams for the survey (Defr\`ere et al.\ in prep). It converts raw NOMIC frames to calibrated null measurements in five main steps: frame selection, background subtraction, stellar flux computation, null computation, and null calibration. Frame selection is done by removing the first 20 frames of each OB that are affected by a transient behavior of the NOMIC detector. Background subtraction is achieved by nod pairs, subtracting from each frame the median of frames recorded in the other nod position. Each row in a given channel is then corrected for low frequency noise \citep{Hoffmann:2014} by subtracting its sigma-clipped median, excluding the region around the star position.  Remaining bad pixels were identified using the local standard deviation (5-$\sigma$ threshold) and replaced with the mean of the neighbor pixels. The flux computation is done in each frame by aperture photometry using an aperture radius of 8 pixels (140\,mas or 2.6\,AU at the distance of $\eta$\,Crv), equivalent to the half width at half maximum (HWHM) of the single-aperture PSF at 11.1\,$\mu$m. The background flux in the photometric aperture is estimated simultaneously in each frame using all pixels covering the same columns as the photometric region and located between the second minimum of the single-aperture PSF (34 pixels or 600 mas) and the channel edge. Raw null measurements are then obtained by dividing the individual flux measurements at null by the total flux estimated from the photometric OB of the same observation (accounting for the 50/50 beamsplitter). This procedure is performed exactly the same way for $\eta$\,Crv and the three calibrators. Finally, in order to correct for the differential background error on the null due to the different brightness of the stars, we further add to the null a small fraction of background flux measured in a nearby empty region of the detector. This fraction is computed to match the background error on the null between the different stars, assuming fully uncorrelated noise in the photometric aperture and the nearby empty region of the detector (see Appendix~\ref{AppA}). To compute the final null value per OB, we first remove the outliers using a 5-$\sigma$ threshold and then perform the weighted-average of the lowest 5\% null measurements (the weight being defined as the inverse square of the uncertainty in the aperture photometry). The corresponding null uncertainty is computed by bootstrapping through the entire dataset (keeping the lowest 5\% each time) and taking the 16 and 84\% levels from the cumulative distribution as representative 1-$\sigma$ uncertainties. A description of the additional systematic errors considered is given in Section~\ref{sec:analysis}. The raw null measurements per OB for $\eta$\,Crv and its calibrators are shown in the top panel of Figure~\ref{fig:TF}. 

The instrumental null floor is estimated using the calibrator null measurements corrected for the finite extension of the stars. This correction is done using a linear limb-darkened model for the geometric stellar null \citep{Absil:2006}:\\
\begin{equation}
N_{\rm star} = \left(\frac{\pi B \theta_{\rm LD}}{4\lambda}\right)^2\left(1-\frac{7u_\lambda}{15}\right)\left(1-\frac{u_\lambda}{3}\right)^{-1};\\
\label{eq:geom}
\end{equation}

\noindent where $B$ is the interferometric baseline, $\theta_{\rm LD}$ is the limb-darkened angular diameter of the photosphere (see Table~\ref{tab:calib}), $\lambda$ is the effective wavelength, and $u_\lambda$ is the linear limb-darkening coefficient. Given the relatively short interferometric baseline of the LBTI, the geometric stellar null will generally be small in the mid-infrared and the effect of limb-darkening negligible. Assuming $u_\lambda$ = 0, the typical geometric null for our targets is  $\sim10^{-5}$ with an error bar of $\sim10^{-6}$, which is small compared to our measurement errors. The next step in the calibration is to compute the instrumental null floor at the time of the science observations and subtract it from the raw null measurements of the science target \citep[null contributions are additive,][]{Hanot:2011}. This can be estimated in various ways, using only bracketting calibrator measurements, a weighted combination of the calibrators, or by polynomial interpolation of all calibrator measurements. In the present case, because $\eta$\,Crv and its calibrators have been chosen close in magnitude and position on the sky, we assume that the instrument behaved consistently for the duration of the observations and use a single/constant to fit the instrumental null floor, as shown by the solid line in the top panel of Figure~\ref{fig:TF}. The only noticeable difference between $\eta$\,Crv (F2V) and its calibrators (G8III to K4III) is the spectral type but the slope of the stellar flux across the N' band is the same so there is no differential chromatic bias on the null measurement.  

The statistical uncertainty on a calibrated null measurement is computed as the quadratic sum of its own statistical uncertainty and the statistical uncertainty on the null floor measured locally at any given time by using the 1-$\sigma$ variation on the constant fit of the instrumental null floor. The systematic uncertainty, on the other hand, is computed globally and is composed of two terms. The first one accounts for the uncertainty on the stellar angular diameters, taking the correlation between calibrators into account. As described above, it is negligible in this case. The second one is computed as the weighted standard deviation of all calibrator measurements and accounts for all other systematic uncertainties, which are harder to estimate on a single-OB basis (see Section~\ref{sec:analysis}). The statistical and systematic uncertainties are uncorrelated and added quadratically to give the total uncertainty on a calibrated null measurement. The part of this uncertainty related to the the null floor is represented by the dashed lines in the top panel of Figure~\ref{fig:TF}. The calibrated null measurements are shown in Section~\ref{sec:discuss}.

\section{Data analysis and results}\label{sec:analysis}

Figure~\ref{fig:TF} shows that the null measurements obtained on the science target (red diamonds) are clearly above the instrumental null floor (blue squares), suggestive of resolved emission around $\eta$\,Crv. In order to quantify the measured excess emission, the first step of the data analysis is to convert the calibrated null measurements into a single value. Given that the calibrated null measurements show no significant variation as a function of the baseline rotation, this is done by computing the weighted average of all calibrated null measurements and gives  $4.40\%  \pm 0.10\% \pm 0.34\%$, where the two uncertainty terms correspond to the statistical and the systematic uncertainties respectively. The statistical uncertainty decreases with the number of data points and is computed as follows:
\begin{equation}
\sigma_{\rm stat} = \frac{1}{\sqrt{\sum_i1/\sigma^2_{i, \rm stat}}};\\
\label{eq:stat}
\end{equation}
%\begin{equation}
%\sigma = \sqrt{\left(\frac{1}{\sqrt{\sum_i1/\sigma_i^{\rm stat}}}\right)^2+\left(\frac{\sum_i\sigma_i^{\rm syst}}{n}\right)^2};\\
%\label{eq:geom}
%\end{equation}
\noindent where $\sigma_{i, \rm stat}$ is the statistical part of the uncertainty on the $i$th calibrated data point. The systematic uncertainty on the other hand has been computed globally and we make here the conservative assumption that it is fully correlated between all data points. The systematic uncertainty on the final calibrated null is hence given by the systematic uncertainty on the null floor, which is the same for all calibrated data points. Several instrumental imperfections contribute to this uncertainty. First, there is  a background measurement bias between the photometric aperture and the nearby regions used for simultaneous background measurement and subtraction. The amplitude of this bias depends on the the photometric stability (i.e., precipitable water vapor, temperature, clouds) and the nodding frequency. It was estimated using various empty regions of the detector located around the photometric aperture. The result is shown for one representative empty region in the bottom panel of Figure~\ref{fig:TF} and accounts for 0.18\% of the systematic uncertainty. Another main contributor to the systematic uncertainty comes from the mean phase setpoint used to track the fringes. Using a more advanced data reduction technique \citep[i.e.,][]{Hanot:2011} on similar data sets, we estimate that this error can produce a null uncertainty as large as 0.2\% between different OBs. Finally, a variable intensity mismatch between the two beams can also impair the null floor stability. In the present case, it was measured in each observation using the photometric OBs and found to be stable at the 1.5\% level, corresponding to a null error of $\sim$0.1\%. Adding quadratically these 3 terms gives a systematic uncertainty of 0.29\%, similar to the value found after data reduction (i.e., 0.34\%).

\begin{figure}[!t]
\centering
\includegraphics[height=8.5cm]{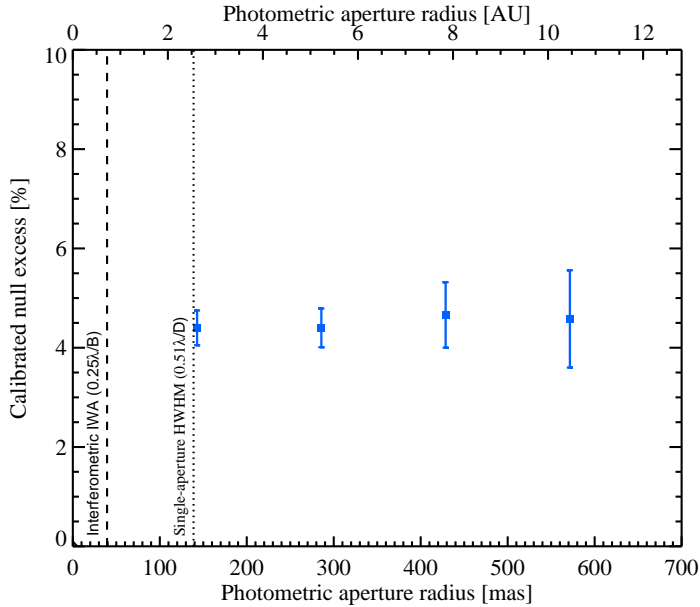}
\caption{Calibrated null excess as a function of photometric aperture radius. The left vertical dashed line indicates the interferometric inner working angle (i.e., $\lambda/4B$=39\,mas or 0.7\,AU at the distance of $\eta$\,Crv) while the right vertical dotted line represents the default aperture radius used in this study (i.e., the half width at half maximum of the single-aperture PSF at 11.1\,$\mu$m). The lack of significant increase in the null measurements for larger aperture radii suggests that the regions emitting the majority of the N'-band flux is more compact than the size of the single-aperture PSF.}
\label{fig:aper}
\end{figure}

The final step of the data analysis is to compute the fraction of the calibrated source null depth that actually comes from the circumstellar environment by subtracting the null depth expected from the stellar photosphere alone. Using a limb-darkened diameter of 0.819 $\pm$ 0.119\,mas (see Table~\ref{tab:calib}) and Equation~\ref{eq:geom}, the stellar contribution to the total null depth is 0.0014\% $\pm$ 0.0004\% which is negligible compared to the calibrated null excess. Therefore, adding quadratically the statistical and systematic uncertainties, the final source null depth detected around $\eta$\,Crv is 4.40\% $\pm$ 0.35\%. To make sure that no dust emission is lying outside the photometric aperture, we reduced the data using larger apertures and following exactly the same procedure. The results are illustrated in Figure~\ref{fig:aper} for four different aperture radii which are multiple of the half width at half maximum of the single-aperture PSF at 11.1\,$\mu$m. The lack of significant increase in the null measurements for larger aperture radii suggests that the angular size of the inner disk is smaller than the size of the single-aperture PSF (2.6\,AU in radius). This result is in agreement with the conclusion from single-dish imaging \citep{Smith:2008} and the KIN \citep{Mennesson:2014}.

%The error bar on the calibrated increases for larger aperture radii because of high-order wavefront null instability and the larger photometric noise. 
%The precipitable water vapor (PWV) was 4.2\,mm and stable as measured by the nearby Submillimeter Telescope (SMT) Tau-meter. 

\section{Modeling and interpretation}\label{sec:model}

A first step toward interpreting the calibrated null depth observed for $\eta$\,Crv can be made simply by comparing it with the ratio of disk to photospheric flux observed with \emph{Spitzer/IRS}.  Using the spectrum presented in \cite{Lebreton:2014} and assuming an absolute calibration error of 3\%, we derive a value of 22.7\% $\pm$ 5.4\% over the N' band. This is a very conservative uncertainty estimate which is consistent with previous studies \citep{Chen:2006,Lisse:2012}. For any centrosymmetric disk, approximately half the disk flux is transmitted through the LBTI fringe pattern unless the extent of the N'-band emission is comparable to, or smaller than the angular resolution of the interferometer (i.e., $\lambda/2B$=79\,mas or 1.4\,AU). The observed null depth of 4.40\% $\pm$ 0.35\%, i.e. substantially lower than the disk to star flux ratio, suggests therefore that the disk is either compact or edge-on with a position angle perpendicular to the LBTI baseline. An edge-on disk perpendicular to the baseline would have a position angle of around 0-20$^\circ$ which is specifically disfavored by VLTI/MIDI observations \citep[see Fig.\ 11 in][]{Smith:2009}. In addition, the inner disk would be nearly perpendicular to the outer disk plane whose orientation has been consistently measured by independent studies \citep[see dashed line in Figure~\ref{fig:TF},][]{Wyatt:2005,Duchene:2014,Lebreton:2014}. Given that the warm dust may be scattered in from the cool outer belt \citep[e.g.,][]{Wyatt:2007,Lisse:2012}, such a configuration seems unlikely and would be difficult to explain dynamically. Besides, among the planetary systems with measured inner and outer disk orientations (e.g., Solar system, $\beta$\,Pic, AU\,Mic), none shows perpendicular inner and outer disks. A more likely scenario is for the inner and outer disk components to be coplanar, and to concentrate more than half the disk flux within 1.4\,AU.

%systems with a disk seen edge-on always appear to be aligned in this way (e.g., the Solar System, beta Pic, AU Mic). %In short, this model assumes that the dust emits at the equilibrium temperature for blackbodies. It is based on a dozen parameters related to the star (e.g., distance, luminosity,..), the disk (e.g., radius, surface density,...), and the observations (inclination and position angle, see complete list in Table\,1 of \citealt[][]{Kennedy:2014}). Each disk parameter has a reference value defined as that of the Solar zodiacal cloud model \citep{Kelsall:1998}. The most relevant ones in the context of our observations are an inner radius of 0.08\,AU (corresponding to the sublimation radius of silicate grains), a reference radius $r_0$ of 2.3\,AU, a surface density of $7.12\times10^{-8}$ at r$_0$, an outer radius of 23\,AU and a density power-law $\propto r^{-\alpha}$ with $\alpha=0.34$. To account for the fact that $\eta$\,Crv is a F2V star, the reference radius is scaled with the square root of the stellar luminosity (i.e., $r_0=\sqrt{L_\star}$=2.3\,AU). The inner and outer radii are also scaled from the Solar system with the square root of the stellar luminosity to maintain the same temperature at the inner and outer edges of the disk as for the Solar zodiacal cloud. Using this reference model, we compute the multiplication factor of the surface density at $r_0$, relative to the Solar model, that would reproduce both the observed null measurement and the disk/star flux ratio measured by \emph{Spitzer/IRS}.

\begin{figure}[!t]
\centering
\includegraphics[height=8.5cm]{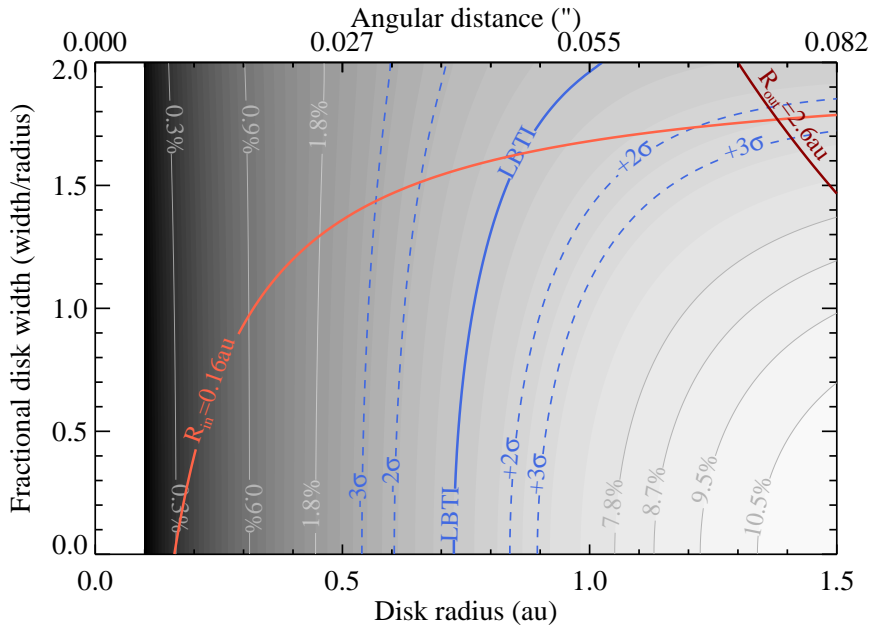}
\caption{Constraints on the location and width of the resolved excess emission detected around $\eta$\,Crv. Grey contours and labels show the predicted LBTI null assuming a disk to star flux ratio of $\sim$23\% and the same orientation as the outer belt. Blue lines show the observed LBTI null and corresponding uncertainties. The left red line shows the inner disk radius lower limit of 0.16\,AU as derived with VLTI/MIDI while the right dark red line indicates the outer limit set by our observations. Considering all constraints, the disk is restricted to lie at a distance of 0.5-1.0\,AU, with a full width of up to 1.7\,AU.}
\label{fig:model}
\end{figure}

In order to interpret our observations within the context of a physical model, we construct a simple vertically thin model for the inner disk using the analytical tool developed by the HOSTS team \citep[][]{Kennedy:2014}. As expected from the qualitative description above, scaled models of the Solar zodiacal cloud that match the observed total and transmitted disk fluxes have position angles near-perpendicular to the outer disk. That is, this model is too radially extended so, unless the inner disk is strongly misaligned with the outer disk, a scaled version of our Solar zodiacal cloud is not a good match to the warm disk around $\eta$\,Crv. We therefore explore the alternative possibility that the inner and outer disks are coplanar. With the disk orientation fixed, it is necessary to vary the location and width of the warm dust disk to reproduce both the total disk flux and the calibrated null measured by LBTI. That is, the only free parameters are the disk radius and width ($r$ and $\delta r$), and the power-law for the dust (i.e., surface density or optical depth) between these radii ($\alpha$). The results of this process, for the radius and width, are shown in Figure~\ref{fig:model}. The greyscale and grey contours show the null depths of models with a range of radii $r$ and fractional widths $\delta r/r$, where in each case the total disk to star flux ratio is 23\% in the N' band (using $\alpha=0.34$). The null depths vary from near zero for small disk radii, to about 11\% for narrow rings with $r \approx 1.5$\,AU. The blue lines show where the predicted null depths agree with the LBTI observations of $\eta$ Crv with the range allowed by 2 and 3-$\sigma$ uncertainties. These uncertainties do not account for the 5\% uncertainty in the \emph{Spitzer/IRS} excess, which for example moves the blue lines systematically approximately 0.1\,AU outward for a 5\% lower excess. Similarly, if the IRS excess is larger, which is possible if the excess is present at shorter wavelengths \citep[e.g.,][]{Lisse:2012}, then the disk lies even closer to the star. With these assumptions, the inner disk component is constrained to lie at approximately 0.5-1.0\,AU with varying width. 

\begin{figure}[!t]
\centering
\includegraphics[height=8.5cm]{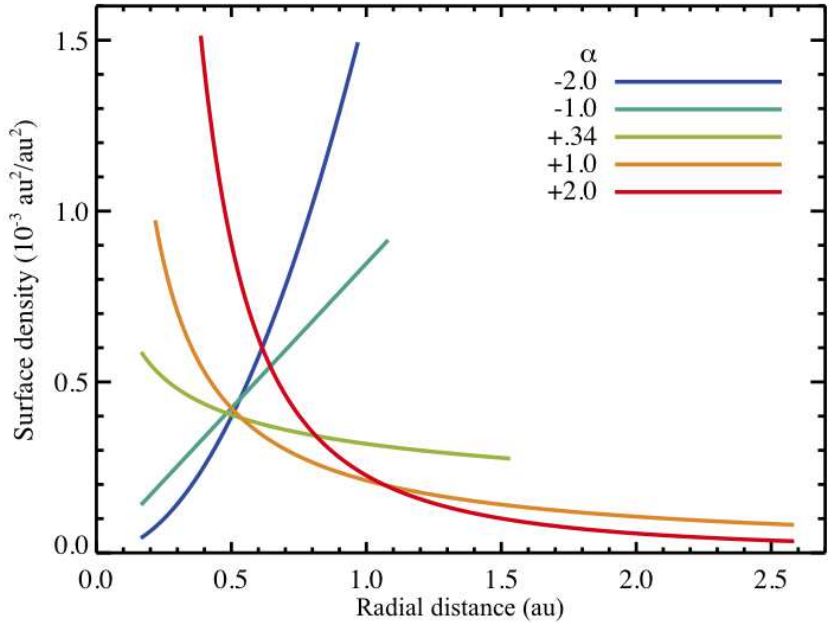}
\caption{Surface density as a function of distance from the star for five disk models having different power-law indices $\alpha$. Each line represents the widest disk allowed in a Figure~\ref{fig:model} computed for the corresponding $\alpha$. For a steeply decreasing surface density profile (i.e., large $\alpha$), the disk can extend beyond 1\,AU but the disk emission is then more concentrated.}
\label{fig:alphas}
\end{figure}

As noted in Section~\ref{sec:analysis}, the lack of increase in the null measurements for larger aperture radii suggests that the disk is smaller than 2.6\,AU in radius. From MIDI, VISIR, and MICHELLE observations, \cite{Smith:2009} constrain the inner disk to lie outside 0.16\,AU. These constraints are shown by the red curves in Figure~\ref{fig:model}, and serve as useful limits if the disk is wide, in which case the LBTI does not constrain the disk extent strongly. The constraint is in fact better than appears from Figure~\ref{fig:model} because even the widest models are concentrated; half of the total disk flux originates from within 0.8\,AU. Wide disks are only allowed because enough of their emission is blocked by the LBTI transmission pattern so the qualitative description given above holds. The zoo of possible models can be made even larger by adjusting the surface density power law. The default value of $\alpha$=0.34 used above comes from the solar Zodiacal cloud model \citep{Kelsall:1998} which is not necessarily a good match to the warm disk around $\eta$\,Crv. It could be very different depending on the physical origin of the disk, and depends on dynamical, collisional, and radiation processes. Given that the warm dust around $\eta$\,Crv may be scattered in from the cool outer belt \citep[e.g.,][]{Wyatt:2007,Lisse:2012}, predictions of the expected $\alpha$ would be highly uncertain. However, its value has a very weak impact on the warm dust location as shown in Figure~\ref{fig:alphas}. For models where the surface density increases with radius ($\alpha<0$), the disk emission is concentrated farther from the star, so the disk must be smaller in order to fit our data. If the disk surface density decreases with radius ($\alpha>0$), our ability to constrain the disk extent is limited. However, for these models most of the emission comes from the inner disk edge, so the inability to constrain the outer edge location is expected, and applies similarly to other observations. Therefore, regardless of the disk profile, the conclusion from any centrosymmetric model is that the bulk of the disk emission lies closer to the star than previously thought, most likely in the range 0.5-1\,AU.

%The adopted fitting stategy is based on a Bayesian inference method. 
%We implement a new method to calculate the sublimation temperature Tsub of a dust grain as a function of its size and not only of its composition

\section{Discussion} \label{sec:discuss}

The warm dust location derived in the previous section is significantly closer to the star than inferred from \emph{Spitzer/IRS} spectrum models \citep[3\,AU or 160\,mas,][]{Lisse:2012}. There is however significant uncertainty in this inference because it relies on complex grain models with degenerate parameters. While the LBTI observations clearly suggest that these models need to be revisited, an alternative scenario to reconcile the difference in the inferred disk sizes is to relax the assumption of a centrosymmetric disk. If the inner disk has an over-density that was hidden behind a transmission minimum for the range of observed hour angles, then the disk can be larger and may still satisfy the observed LBTI null and \emph{Spitzer/IRS} excess (i.e., a clump can be at larger physical separation, but remain close to the star in sky-projected distance). Future observations can therefore aim to expand the range of hour angles and also look for significant variations in the calibrated null depth at the same hour angles. Such observations will help establish whether an asymmetry exists, and if so, whether that asymmetry is orbiting the star or fixed in space. The latter possibility would favor recent  collision scenarios which suggest that a long-lasting asymmetry is created at the location where the impact occurred \citep[][]{Jackson:2014}.

%While blackbody models of the inner disk find a temperature of 350-400\,K, which places the disk near 1\,AU \citep[e.g.,][]{Wyatt:2005,Smith:2008}, the likely %presence of small grains that are warmer than blackbodies for the same stellocentric distance means that the inner disk is likely to be nearer to 3\,AU %%%\citep[160\,mas,][]{Lisse:2012}. 

\begin{figure}[!t]
\centering
\includegraphics[height=8.5cm]{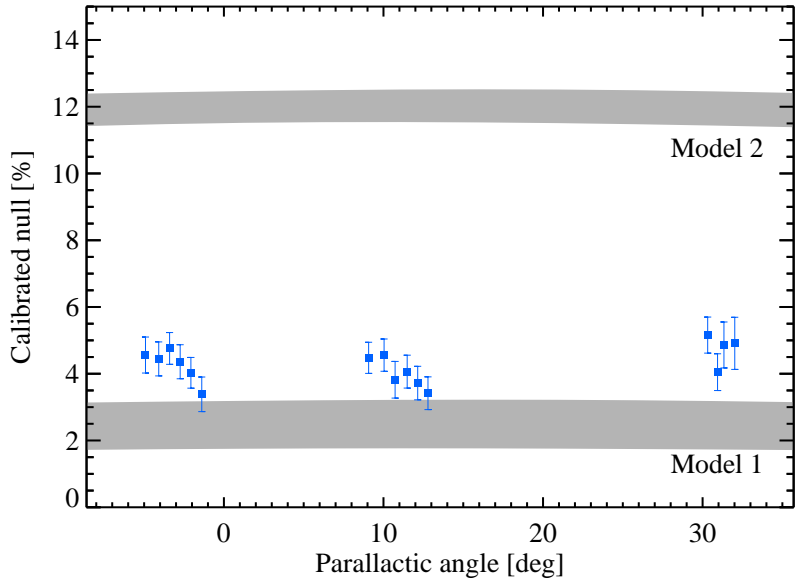}
\caption{Calibrated null measurements as a function of parallactic angle for our LBTI observations of $\eta$\,Crv (February 12, 2014). The null expected from the two best-fit exozodiacal disk models derived by \citet{Lebreton:2014} to reconcile the KIN observations with the shape of the \emph{Spitzer/IRS} spectrum are shown by the grey-shaded areas (see main text for model description). The width of the grey-shaded areas is set by the uncertainty on the density peak location of the model.}
\label{fig:model_vs_null}
\end{figure}

\begin{figure}[!ht]
\centering
\includegraphics[width=6.6cm]{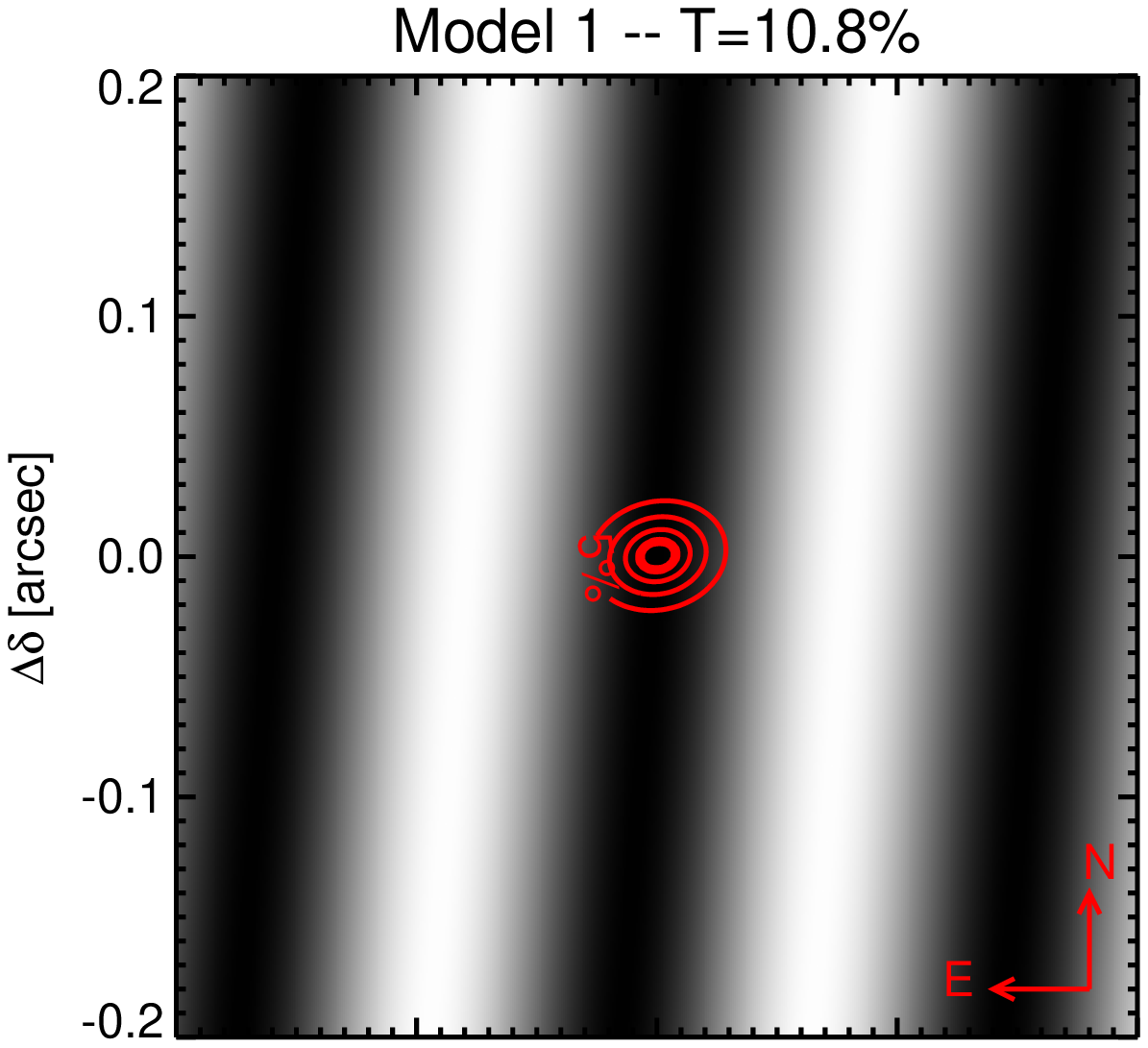}
\includegraphics[width=6.6cm]{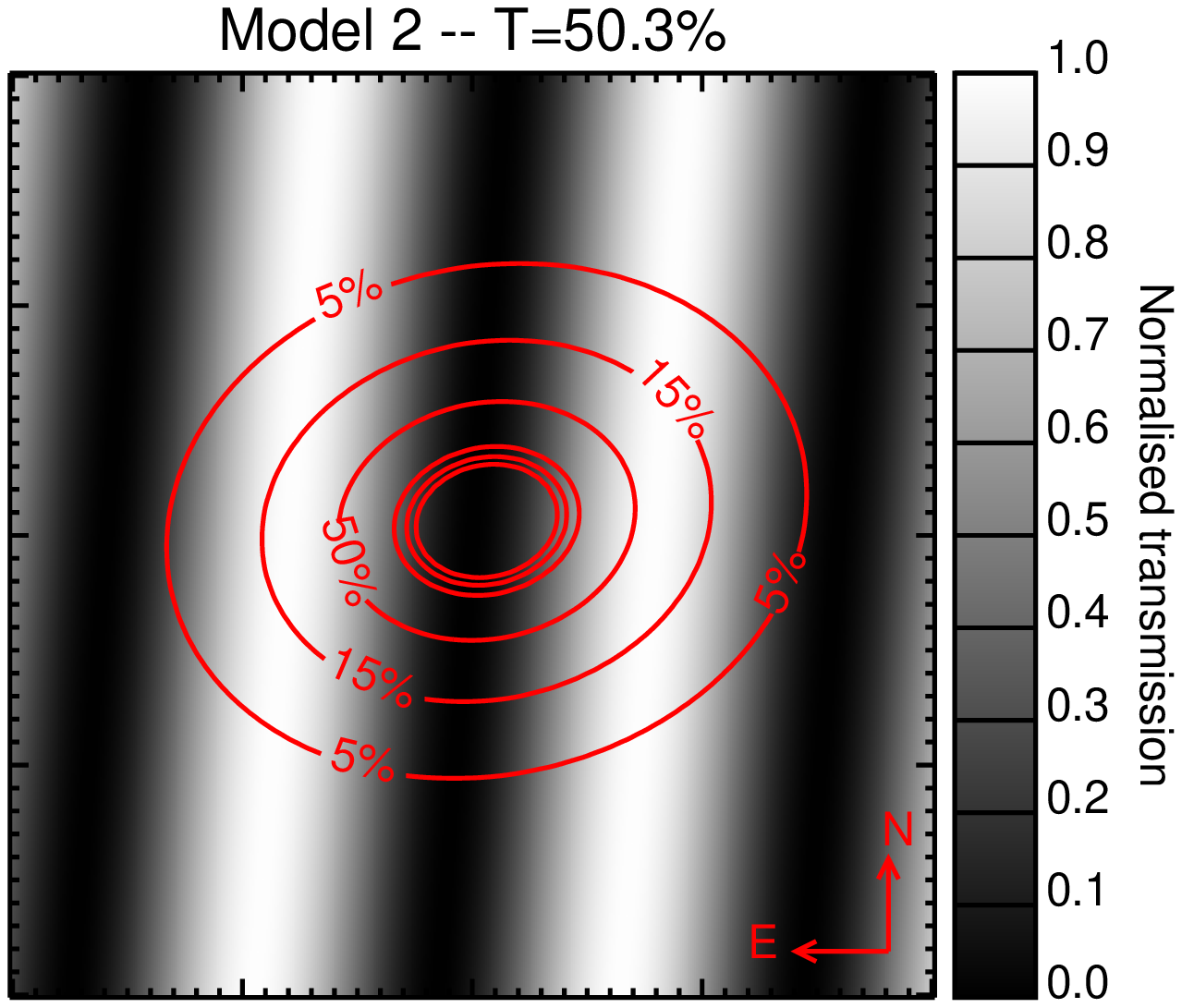}
\includegraphics[width=6.6cm]{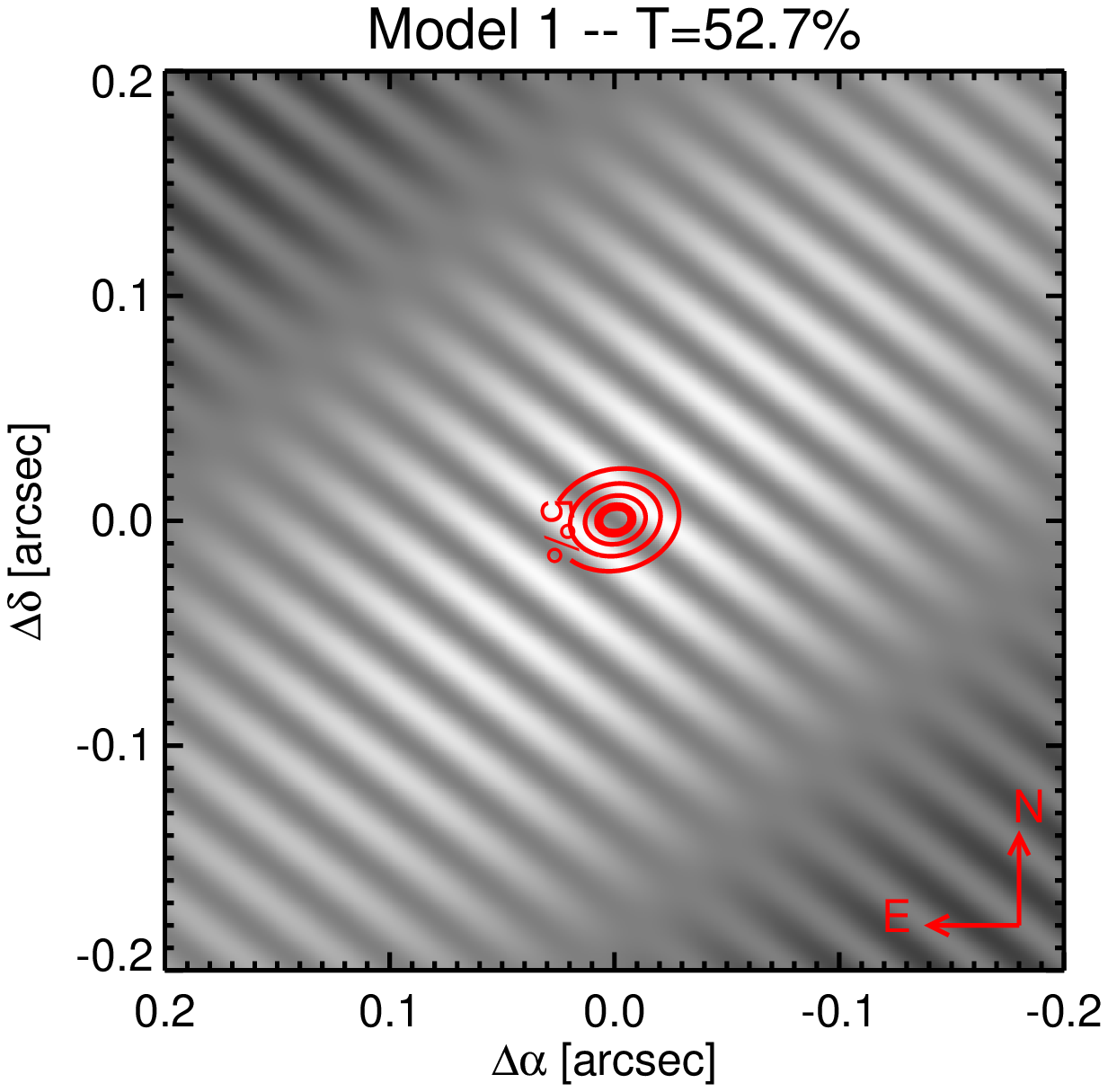}
\includegraphics[width=6.6cm]{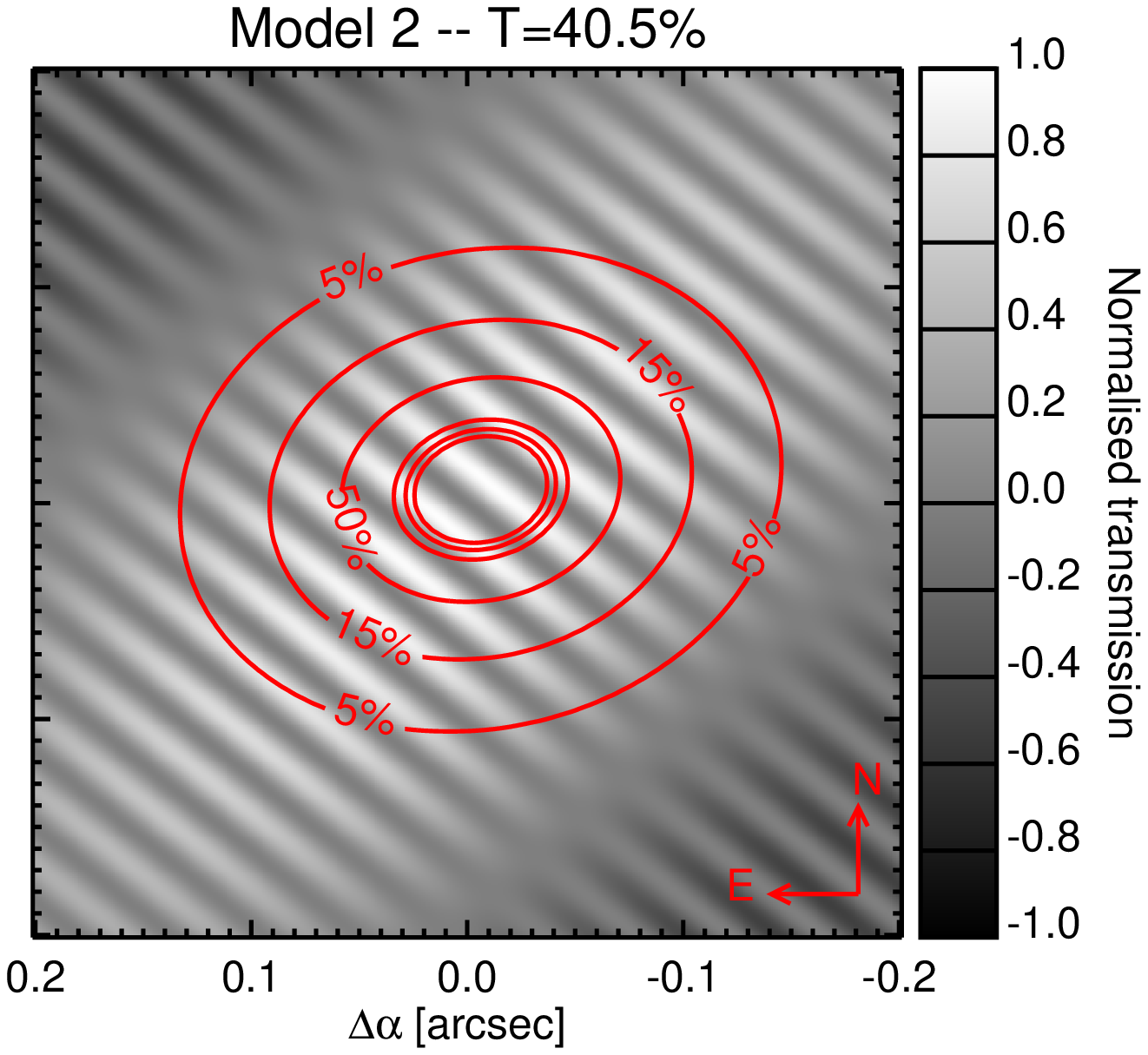}
\caption{The two best-fit exozodiacal disk models derived by \citet{Lebreton:2014} to reconcile the KIN observations with the shape of the \emph{Spitzer/IRS} spectrum shown on top of the sky transmission map of the LBTI (top row) and the KIN (bottom row). Model 1 is on the left and model 2 on the right (see main text for model descriptions). The contours mark regions corresponding to given ratios of the maximum disk flux density (see contour label), assuming that the inner warm disk has the same orientation as the outer disk imaged by \emph{Herschel} \citep[][]{Duchene:2014}. While the predicted flux transmitted through the KIN is logically similar for both models (see value in the upper left legend), it is significantly different in the case of the LBTI which is well matched for few AU-scale dust emission around near main-sequence stars and allows to break the degeneracy between the models (see Figure~\ref{fig:model_vs_null} for the corresponding expected null measurements). The LBTI sky transmission map is shown for the first OB of our observations (Julian date: 2456700.9204, parallactic angle: -4\fdg9, hour angle: -0h18). The KIN sky transmission map is shown for the same hour angle. North is up, East is to the left. }
\label{fig:model2}
\end{figure}

New modeling based on \emph{Spitzer/IRS} models and KIN observations also suggests a compact warm disk emitting mostly from within 1\,AU \citep{Lebreton:2014}. Using the \emph{GraTer} radiative transfer modeling code \citep{Augereau:1999}, \citet{Lebreton:2014} propose two inner disk models that reconcile the KIN observations with the shape of the mid-infrared \emph{Spitzer/IRS} spectrum and various photometric constraints from \emph{IRAS}, \emph{AKARI}, \emph{WISE}, and \emph{Spitzer/MIPS}. The first model (model 1 hereafter) consists of small amorphous forsterite \citep[Mg$_2$SiO$_4$,][]{Jager:2003} dust grains with a differential size distribution defined by $n(a) \propto a^{-\kappa}$, with $\kappa=4.5$ and $a_{\rm min}=1.2~{\micron}$. The density peak location of the grains is 0.16-0.25\,AU with a radial decrease in surface density $\propto r^{-1.5}$. Note that forsterite grains are also the dominant species in the \emph{Spitzer/IRS} best-fit model of \cite{Lisse:2012}. The second model (model 2 hereafter) consists of astronomical silicates \citep{Draine:1984}, with a minimum size of $a_{\rm min}=0.6~{\micron}$ and a size distribution slope of $\kappa=3.5$. The density peak location of the grains is 0.73-0.85\,AU with an outer density slope of $-1$. The expected null excess for both models is shown in Figure~\ref{fig:model_vs_null} as a function of parallactic angle. As indicated by the grey-shaded areas, model 1 is a better fit to the LBTI observations, particularly for its larger value of the density peak position (0.25\,AU). To understand how the LBTI observations break the model degeneracy, the two disk models are shown on top of the LBTI and KIN transmission maps in Figure~\ref{fig:model2}. While both models logically produce a similar flux at the output of the KIN, the expected flux transmitted to the null output of the LBTI is significantly lower for model 1 due to the shorter nulling baseline and the small extent of the disk model. Approximately 10-19\% of the disk flux is transmitted to the null output for model 1, depending on the density peak position (i.e., 0.16-0.25\,AU), while this value reaches 48-51\% for model 2 (0.73-0.85\,AU). This explains the difference in the predicted null excess shown in Figure~\ref{fig:model_vs_null}. While the parameters of model 1 could be tuned to better fit our observations, this analysis clearly shows that it is possible to reconcile the \emph{Spitzer/IRS} model with spatial constraints from the KIN and the LBTI. The very steep size distribution and resulting large population of very small grains of such a model depart significantly from the behavior expected for a collisional cascade in equilibrium \citep{Gaspar:2012}. It is possible that this is evidence for the dust being the product of a transient, high rate of collisions \citep[e.g.,][]{Olofsson:2012}.  

An important question related to the overall goal of LBTI is what the dust level in the habitable zone around $\eta$\,Crv is. From the above discussion, it is clear that a scaled version of the Solar zodiacal dust cloud is not a good match to the observations and quoting a ``zodi'' level could therefore be confusing \citep[see][for a discussion about this unit]{Roberge:2012}. To quantify the amount of warm dust, we can instead compare the surface density of our models with that of the Solar zodiacal cloud at the same equivalent position (i.e., scaled as $\sqrt{L_\star}$). Several possible scenarios have been discussed, and each will make different predictions. In the simplest scenario, where the inner disk is coplanar with the outer disk, the analytical model presented in Figure~\ref{fig:model} has a surface density at 0.7\,AU approximately 10$^4$ times greater than that at the equivalent location in the Solar zodiacal cloud (i.e., 0.3\,AU). At 2.3\,AU, the distance where the insolation is the same as Earth's at $\sqrt{L_\star}$, it has formally no dust since models within the 2-$\sigma$ contours in Figure~\ref{fig:model} all have outer edges within 2\,AU. The radiative transfer disk model (model 1) on the other hand is not truncated at its outer edge (surface density $\propto r^{-1.5}$) and has a surface density at 2.3\,AU approximately 12000 times larger than that at 1\,AU in the Solar zodiacal cloud. Finally, in the non-centrosymmetric scenarios, the possibility of even more dust at larger radii means that the habitable zone dust levels could be even larger, but localized to certain azimuthal ranges which would be a source of confusion for future imaging missions \citep[e.g.,][]{Defrere:2012}. Even if the lack of variation of the null measurements as a function of the observed baseline position angle suggests that the disk is most likely centrosymmetric, this scenario cannot be ruled out with the present data and more observations expanding the range of sky rotation are required.  

%It is already clear that this star is not a good target for an exo-Earth imaging mission. 
%Similarly, the aperture analysis finds that the bulk of the dust emission lies within 2.6AU, but the uncertainties on the calibrated null depth mean that the limits on dust beyond this radius are still MAYBE of order 1-2 %orders of magnitude above the Solar zodiacal level. 

\section{Summary and conclusions} 

This paper presents the first science results from the LBTI nuller, a mid-infrared interferometric instrument combining the two 8.4-m apertures of the LBT. As part of LBTI's commissioning observations, we observed the nearby main-sequence star $\eta$\,Crv, previously known to harbor high levels of circumstellar dust. The data are consistent with a source null depth of 4.40\% $\pm$ 0.35\% across the N' band (9.81 - 12.41\,$\mu$m) and over a field-of-view of 140\,mas in radius  ($\sim$2.6\,AU at the distance of $\eta$\,Crv). The measured null shows no significant variation over 35$^\circ$ of sky rotation and is relatively low compared to the total disk to star flux ratio of 23\% observed with \emph{Spitzer/IRS}, suggesting that a significant fraction of the dust lies within the central nulled response of the LBTI (79\,mas or 1.4\,AU along the baseline orientation). Assuming that the inner and outer disk components are coplanar, we show that the bulk of the warm dust emission must lie at a distance of 0.5-1.0\,AU from the star in order to fit our data. This is consistent with spatial constraints from the KIN that also point toward a compact warm disk emitting mostly from within 1\,AU but significantly closer than the distance predicted by the \emph{Spitzer/IRS} spectrum models (3\,AU). This discrepancy illustrates how spatially resolved observations provide crucial information to break the degeneracy inherent to SED-based modeling of proto-planetary and debris disks showing evidence of warm emission. This is discussed in detail for $\eta$\,Crv in a companion paper based on the KIN observations \citep{Lebreton:2014}. In this paper, we show how the LBTI observations break the degeneracy between models that have the same SEDs and KIN nulls and discuss an alternative scenario which is the possible presence of an over-density in the inner disk. Both scenarios support the prevailing interpretation for the origin of the warm dust which is a recent collision in the inner planetary system. To conclude, it is clear that $\eta$\,Crv is not a good target for a future exo-Earth imaging mission since the surface density of the dust in or near the habitable zone can be as high as four orders of magnitude larger than that in the Solar zodiacal cloud.
%Although attractive, this scenario is less likely given the lack of variation of the null measurements as a function of the observed parallactic angles (see Figure~\ref{fig:model_vs_null}). 
%The results presented in this paper were obtained as part of LBTI's commissioning science operations and are not representative of the ultimate performance that will be delivered by the LBTI. 
% A more extended edge-on disk perpendicular to the baseline could fit the data but would imply that the disk is strongly misaligned with the outer disk imaged by \emph{Herschel} which is hard to explain dynamically

\begin{acknowledgements}
The Large Binocular Telescope Interferometer is funded by the National Aeronautics and Space Administration as part of its Exoplanet Exploration Program. The LBT is an international collaboration among institutions in the United States, Italy and Germany. LBT Corporation partners are: The University of Arizona on behalf of the Arizona university system; Istituto Nazionale di Astrofisica, Italy; LBT Beteiligungsgesellschaft, Germany, representing the Max-Planck Society, the Astrophysical Institute Potsdam, and Heidelberg University; The Ohio State University, and The Research Corporation, on behalf of The University of Notre Dame, University of Minnesota and University of Virginia. This work was supported by the European Union through ERC grant number 279973 (GMK \& MCW). The authors thank C. Lisse for helpful advice. The LBTI team would like to dedicate this paper to the memory of our colleague and friend, Vidhya Vaitheeswaran.
\end{acknowledgements}

\bibliographystyle{apj}

\begin{thebibliography}{}
\expandafter\ifx\csname natexlab\endcsname\relax\def\natexlab#1{#1}\fi

\bibitem[{{Absil} {et~al.}(2006){Absil}, {den Hartog}, {Gondoin}, {Fabry},
  {Wilhelm}, {Gitton}, \& {Puech}}]{Absil:2006}
{Absil}, O., {den Hartog}, R., {Gondoin}, P., {et~al.} 2006, \aap, 448, 787

\bibitem[{{Absil} {et~al.}(2013){Absil}, {Defr{\`e}re}, {Coud{\'e} du Foresto},
  {Di Folco}, {M{\'e}rand}, {Augereau}, {Ertel}, {Hanot}, {Kervella},
  {Mollier}, {Scott}, {Che}, {Monnier}, {Thureau}, {Tuthill}, {ten Brummelaar},
  {McAlister}, {Sturmann}, {Sturmann}, \& {Turner}}]{Absil:2013}
{Absil}, O., {Defr{\`e}re}, D., {Coud{\'e} du Foresto}, V., {et~al.} 2013,
  \aap, 555, A104

\bibitem[{{Augereau} {et~al.}(1999){Augereau}, {Lagrange}, {Mouillet},
  {Papaloizou}, \& {Grorod}}]{Augereau:1999}
{Augereau}, J.~C., {Lagrange}, A.~M., {Mouillet}, D., {Papaloizou}, J.~C.~B.,
  \& {Grorod}, P.~A. 1999, \aap, 348, 557

\bibitem[{{Bailey} {et~al.}(2014){Bailey}, {Hinz}, {Puglisi}, {Esposito},
  {Vaitheeswaran}, {Skemer}, {Defr{\`e}re}, {Vaz}, \&
  {Leisenring}}]{Bailey:2014}
{Bailey}, V.~P., {Hinz}, P.~M., {Puglisi}, A.~T., {et~al.} 2014, in Society of
  Photo-Optical Instrumentation Engineers (SPIE) Conference Series, Vol. 9148,
  Society of Photo-Optical Instrumentation Engineers (SPIE) Conference Series,
  3

\bibitem[{{Beichman} {et~al.}(2006{\natexlab{a}}){Beichman}, {Lawson}, {Lay},
  {Ahmed}, {Unwin}, \& {Johnston}}]{Beichman:2006}
{Beichman}, C., {Lawson}, P., {Lay}, O., {et~al.} 2006{\natexlab{a}}, in Proc.
  SPIE, Vol. 6268

\bibitem[{{Beichman} {et~al.}(2006{\natexlab{b}}){Beichman}, {Bryden},
  {Stapelfeldt}, {Gautier}, {Grogan}, {Shao}, {Velusamy}, {Lawler}, {Blaylock},
  {Rieke}, {Lunine}, {Fischer}, {Marcy}, {Greaves}, {Wyatt}, {Holland}, \&
  {Dent}}]{Beichman:2006b}
{Beichman}, C.~A., {Bryden}, G., {Stapelfeldt}, K.~R., {et~al.}
  2006{\natexlab{b}}, \apj, 652, 1674

\bibitem[{{Bonneau} {et~al.}(2011){Bonneau}, {Delfosse}, {Mourard}, {Lafrasse},
  {Mella}, {Cetre}, {Clausse}, \& {Zins}}]{Bonneau:2011}
{Bonneau}, D., {Delfosse}, X., {Mourard}, D., {et~al.} 2011, \aap, 535, A53

\bibitem[{{Bonsor} {et~al.}(2012){Bonsor}, {Augereau}, \&
  {Th{\'e}bault}}]{Bonsor:2012}
{Bonsor}, A., {Augereau}, J.-C., \& {Th{\'e}bault}, P. 2012, \aap, 548, A104

\bibitem[{{Bracewell}(1978)}]{Bracewell:1978}
{Bracewell}, R.~N. 1978, \nat, 274, 780

\bibitem[{{Chen} {et~al.}(2006){Chen}, {Sargent}, {Bohac}, {Kim},
  {Leibensperger}, {Jura}, {Najita}, {Forrest}, {Watson}, {Sloan}, \&
  {Keller}}]{Chen:2006}
{Chen}, C.~H., {Sargent}, B.~A., {Bohac}, C., {et~al.} 2006, \apjs, 166, 351

\bibitem[{{Danchi} {et~al.}(2014){Danchi}, {Bailey}, {Bryden}, {Defr{\`e}re},
  {Haniff}, {Hinz}, {Kennedy}, {Mennesson}, {Millan-Gabet}, {Rieke}, {Roberge},
  {Serabyn}, {Skemer}, {Stapelfeldt}, {Weinberger}, \& M.}]{Danchi:2014}
{Danchi}, W., {Bailey}, V., {Bryden}, G., {et~al.} 2014, in Proc. SPIE, Vol.
  9146

\bibitem[{{Defr{\`e}re} {et~al.}(2010){Defr{\`e}re}, {Absil}, {den Hartog},
  {Hanot}, \& {Stark}}]{Defrere:2010}
{Defr{\`e}re}, D., {Absil}, O., {den Hartog}, R., {Hanot}, C., \& {Stark}, C.
  2010, \aap, 509, A9+

\bibitem[{{Defr{\`e}re} {et~al.}(2012){Defr{\`e}re}, {Stark}, {Cahoy}, \&
  {Beerer}}]{Defrere:2012}
{Defr{\`e}re}, D., {Stark}, C., {Cahoy}, K., \& {Beerer}, I. 2012, in Society
  of Photo-Optical Instrumentation Engineers (SPIE) Conference Series, Vol.
  8442, Society of Photo-Optical Instrumentation Engineers (SPIE) Conference
  Series, 0

\bibitem[{{Defr{\`e}re} {et~al.}(2014){Defr{\`e}re}, Hinz, Downey, Ashby,
  Bailey, Brusa, Christou, Danchi, Grenz, Hill, Hoffmann, Leisenring, Lozi,
  McMahon, Mennesson, Millan-Gabet, Montoya, Powell, Skemer, Vaitheeswaran,
  Vaz, \& Veillet}]{Defrere:2014}
{Defr{\`e}re}, D., Hinz, P., Downey, E., {et~al.} 2014, in , 914609--914609--8

\bibitem[{{Draine} \& {Lee}(1984)}]{Draine:1984}
{Draine}, B.~T., \& {Lee}, H.~M. 1984, \apj, 285, 89

\bibitem[{{Duch{\^e}ne} {et~al.}(2014){Duch{\^e}ne}, {Arriaga}, {Wyatt},
  {Kennedy}, {Sibthorpe}, {Lisse}, {Holland}, {Wisniewski}, {Clampin}, {Kalas},
  {Pinte}, {Wilner}, {Booth}, {Horner}, {Matthews}, \&
  {Greaves}}]{Duchene:2014}
{Duch{\^e}ne}, G., {Arriaga}, P., {Wyatt}, M., {et~al.} 2014, \apj, 784, 148

\bibitem[{{Esposito} {et~al.}(2010){Esposito}, {Riccardi},
  {Quir{\'o}s-Pacheco}, {Pinna}, {Puglisi}, {Xompero}, {Briguglio}, {Busoni},
  {Fini}, {Stefanini}, {Brusa}, {Tozzi}, {Ranfagni}, {Pieralli}, {Guerra},
  {Arcidiacono}, \& {Salinari}}]{Esposito:2010}
{Esposito}, S., {Riccardi}, A., {Quir{\'o}s-Pacheco}, F., {et~al.} 2010, \ao,
  49, G174

\bibitem[{{Esposito} {et~al.}(2012){Esposito}, {Riccardi}, {Pinna}, {Puglisi},
  {Quir{\'o}s-Pacheco}, {Arcidiacono}, {Xompero}, {Briguglio}, {Busoni},
  {Fini}, {Argomedo}, {Gherardi}, {Agapito}, {Brusa}, {Miller}, {Guerra Ramon},
  {Boutsia}, \& {Stefanini}}]{Esposito:2012}
{Esposito}, S., {Riccardi}, A., {Pinna}, E., {et~al.} 2012, in Society of
  Photo-Optical Instrumentation Engineers (SPIE) Conference Series, Vol. 8447,
  Society of Photo-Optical Instrumentation Engineers (SPIE) Conference Series

\bibitem[{{G{\'a}sp{\'a}r} {et~al.}(2012){G{\'a}sp{\'a}r}, {Psaltis}, {Rieke},
  \& {{\"O}zel}}]{Gaspar:2012}
{G{\'a}sp{\'a}r}, A., {Psaltis}, D., {Rieke}, G.~H., \& {{\"O}zel}, F. 2012,
  \apj, 754, 74

\bibitem[{{G{\'a}sp{\'a}r} {et~al.}(2013){G{\'a}sp{\'a}r}, {Rieke}, \&
  {Balog}}]{Gaspar:2013}
{G{\'a}sp{\'a}r}, A., {Rieke}, G.~H., \& {Balog}, Z. 2013, \apj, 768, 25

\bibitem[{{Gomes} {et~al.}(2005){Gomes}, {Levison}, {Tsiganis}, \&
  {Morbidelli}}]{Gomes:2005}
{Gomes}, R., {Levison}, H.~F., {Tsiganis}, K., \& {Morbidelli}, A. 2005, \nat,
  435, 466

\bibitem[{{Hanot} {et~al.}(2011){Hanot}, {Mennesson}, {Martin}, {Liewer},
  {Loya}, {Mawet}, {Riaud}, {Absil}, \& {Serabyn}}]{Hanot:2011}
{Hanot}, C., {Mennesson}, B., {Martin}, S., {et~al.} 2011, \apj, 729, 110

\bibitem[{{Hill} {et~al.}(2014){Hill}, {Ashby}, {Brynnel}, {Christou},
  {Little}, {Summers}, {Veillet}, \& {Wagner}}]{Hill:2014}
{Hill}, J., {Ashby}, D., {Brynnel}, J., {et~al.} 2014, in Proc. SPIE, Vol. 9145

\bibitem[{{Hinz}(2013)}]{Hinz:2013}
{Hinz}, P. 2013, in American Astronomical Society Meeting Abstracts, Vol. 221,
  American Astronomical Society Meeting Abstracts 221, 403.06

\bibitem[{{Hinz} {et~al.}(2012){Hinz}, {Arbo}, {Bailey}, {Connors}, {Durney},
  {Esposito}, {Hoffmann}, {Jones}, {Leisenring}, {Montoya}, {Nash}, {Nelson},
  {McMahon}, {Pinna}, {Puglisi}, {Skemer}, {Skrutskie}, \&
  {Vaitheeswaran}}]{Hinz:2012}
{Hinz}, P., {Arbo}, P., {Bailey}, V., {et~al.} 2012, in Society of
  Photo-Optical Instrumentation Engineers (SPIE) Conference Series, Vol. 8445,
  Society of Photo-Optical Instrumentation Engineers (SPIE) Conference Series

\bibitem[{{Hinz} {et~al.}(2000){Hinz}, {Angel}, {Woolf}, {Hoffmann}, \&
  {McCarthy}}]{Hinz:2000}
{Hinz}, P.~M., {Angel}, J.~R.~P., {Woolf}, N.~J., {Hoffmann}, W.~F., \&
  {McCarthy}, D.~W. 2000, in Proc. SPIE, ed. P.~J. {Lena} \& A.~{Quirrenbach},
  Vol. 4006, 349--353

\bibitem[{{Hoffmann} {et~al.}(2014){Hoffmann}, {Hinz}, {Defr{\`e}re},
  {Leisenring}, {Skemer}, {Arbo}, {Montoya}, \& {Mennesson}}]{Hoffmann:2014}
{Hoffmann}, W.~F., {Hinz}, P.~M., {Defr{\`e}re}, D., {et~al.} 2014, in Society
  of Photo-Optical Instrumentation Engineers (SPIE) Conference Series, Vol.
  9147, Society of Photo-Optical Instrumentation Engineers (SPIE) Conference
  Series, 1

\bibitem[{{Ibukiyama} \& {Arimoto}(2002)}]{Ibukiyama:2002}
{Ibukiyama}, A., \& {Arimoto}, N. 2002, \aap, 394, 927

\bibitem[{{Jackson} {et~al.}(2014){Jackson}, {Wyatt}, {Bonsor}, \&
  {Veras}}]{Jackson:2014}
{Jackson}, A.~P., {Wyatt}, M.~C., {Bonsor}, A., \& {Veras}, D. 2014, \mnras,
  440, 3757

\bibitem[{{J{\"a}ger} {et~al.}(2003){J{\"a}ger}, {Il'in}, {Henning},
  {Mutschke}, {Fabian}, {Semenov}, \& {Voshchinnikov}}]{Jager:2003}
{J{\"a}ger}, C., {Il'in}, V.~B., {Henning}, T., {et~al.} 2003, \jqsrt, 79, 765

\bibitem[{{Kelsall} {et~al.}(1998){Kelsall}, {Weiland}, {Franz}, {Reach},
  {Arendt}, {Dwek}, {Freudenreich}, {Hauser}, {Moseley}, {Odegard},
  {Silverberg}, \& {Wright}}]{Kelsall:1998}
{Kelsall}, T., {Weiland}, J.~L., {Franz}, B.~A., {et~al.} 1998, \apj, 508, 44

\bibitem[{{Kennedy} {et~al.}(submitted to ApJ){Kennedy}, {Wyatt}, {Bryden},
  {Danchi}, {Defrere}, {Defrere}, {Defrere}, {Defrere}, {Defrere}, {Defrere},
  \& {Defrere}}]{Kennedy:2014}
{Kennedy}, G., {Wyatt}, M., {Bryden}, G., {et~al.} submitted to ApJ

\bibitem[{{Kennedy} \& {Wyatt}(2013)}]{Kennedy:2013}
{Kennedy}, G.~M., \& {Wyatt}, M.~C. 2013, \mnras, 433, 2334

\bibitem[{{Lawler} {et~al.}(2009){Lawler}, {Beichman}, {Bryden}, {Ciardi},
  {Tanner}, {Su}, {Stapelfeldt}, {Lisse}, \& {Harker}}]{Lawler:2009}
{Lawler}, S.~M., {Beichman}, C.~A., {Bryden}, G., {et~al.} 2009, \apj, 705, 89

\bibitem[{{Lebreton} {et~al.}(in prep){Lebreton}, {Beichman}, {Bryden},
  {Mennesson}, {Millan-Gabet}, {Defr{\`e}re}, \& {Hinz}}]{Lebreton:2014}
{Lebreton}, J., {Beichman}, C., {Bryden}, G., {et~al.} in prep

\bibitem[{{Lisse} {et~al.}(2012){Lisse}, {Wyatt}, {Chen}, {Morlok}, {Watson},
  {Manoj}, {Sheehan}, {Currie}, {Thebault}, \& {Sitko}}]{Lisse:2012}
{Lisse}, C.~M., {Wyatt}, M.~C., {Chen}, C.~H., {et~al.} 2012, \apj, 747, 93

\bibitem[{{Mallik} {et~al.}(2003){Mallik}, {Parthasarathy}, \&
  {Pati}}]{Mallik:2003}
{Mallik}, S.~V., {Parthasarathy}, M., \& {Pati}, A.~K. 2003, \aap, 409, 251

\bibitem[{{Mennesson} {et~al.}(in press){Mennesson}, {Millan-Gabet}, {Serabyn},
  {Colavita}, {Absil}, {Bryden}, {Wyatt}, {Danchi}, {Defr\`{e}re}, {Dore},
  {Hinz}, {Kuchner}, {Ragland}, {Scott}, {Stapeldfeldt}, {Traub}, \&
  {Woilliez}}]{Mennesson:2014}
{Mennesson}, B., {Millan-Gabet}, R., {Serabyn}, E., {et~al.} in press, \apj

\bibitem[{{M{\'e}rand} {et~al.}(2005){M{\'e}rand}, {Bord{\'e}}, \& {Coud{\'e}
  Du Foresto}}]{Merand:2005}
{M{\'e}rand}, A., {Bord{\'e}}, P., \& {Coud{\'e} Du Foresto}, V. 2005, \aap,
  433, 1155

\bibitem[{{Millan-Gabet} {et~al.}(2011){Millan-Gabet}, {Serabyn}, {Mennesson},
  {Traub}, {Barry}, {Danchi}, {Kuchner}, {Stark}, {Ragland}, {Hrynevych},
  {Woillez}, {Stapelfeldt}, {Bryden}, {Colavita}, \&
  {Booth}}]{Millan-gabet:2011}
{Millan-Gabet}, R., {Serabyn}, E., {Mennesson}, B., {et~al.} 2011, \apj, 734,
  67

\bibitem[{{Olofsson} {et~al.}(2012){Olofsson}, {Juh{\'a}sz}, {Henning},
  {Mutschke}, {Tamanai}, {Mo{\'o}r}, \& {{\'A}brah{\'a}m}}]{Olofsson:2012}
{Olofsson}, J., {Juh{\'a}sz}, A., {Henning}, T., {et~al.} 2012, \aap, 542, A90

\bibitem[{{Roberge} {et~al.}(2012){Roberge}, {Chen}, {Millan-Gabet},
  {Weinberger}, {Hinz}, {Stapelfeldt}, {Absil}, {Kuchner}, \&
  {Bryden}}]{Roberge:2012}
{Roberge}, A., {Chen}, C.~H., {Millan-Gabet}, R., {et~al.} 2012, \pasp, 124,
  799

\bibitem[{Skemer {et~al.}(2014)Skemer, Hinz, Esposito, Skrutskie,
  {Defr{\`e}re}, Bailey, Leisenring, Apai, Biller, Bonnefoy, Brandner, Buenzli,
  Close, Crepp, De~Rosa, Desidera, Eisner, Fortney, Henning, Hofmann, Kopytova,
  Maire, Males, Millan-Gabet, Morzinski, Oza, Patience, Rajan, Rieke, Schertl,
  Schlieder, Su, Vaz, Ward-Duong, Weigelt, Woodward, \&
  Zimmerman}]{Skemer:2014}
Skemer, A.~J., Hinz, P., Esposito, S., {et~al.} 2014, High contrast imaging at
  the LBT: the LEECH exoplanet imaging survey, doi:10.1117/12.2057277

\bibitem[{{Smith} {et~al.}(2008){Smith}, {Wyatt}, \& {Dent}}]{Smith:2008}
{Smith}, R., {Wyatt}, M.~C., \& {Dent}, W.~R.~F. 2008, \aap, 485, 897

\bibitem[{{Smith} {et~al.}(2009){Smith}, {Wyatt}, \& {Haniff}}]{Smith:2009}
{Smith}, R., {Wyatt}, M.~C., \& {Haniff}, C.~A. 2009, \aap, 503, 265

\bibitem[{{Song} {et~al.}(2005){Song}, {Zuckerman}, {Weinberger}, \&
  {Becklin}}]{Song:2005}
{Song}, I., {Zuckerman}, B., {Weinberger}, A.~J., \& {Becklin}, E.~E. 2005,
  \nat, 436, 363

\bibitem[{{Stencel} \& {Backman}(1991)}]{Stencel:1991}
{Stencel}, R.~E., \& {Backman}, D.~E. 1991, \apjs, 75, 905

\bibitem[{{Veillet} {et~al.}(2014){Veillet}, {Brynnel}, {Hill}, {Wagner},
  {Ashby}, {Christou}, {Little}, \& {Summers}}]{Veillet:2014}
{Veillet}, C., {Brynnel}, J., {Hill}, J., {et~al.} 2014, in Proc. SPIE, Vol.
  9149

\bibitem[{{Vican}(2012)}]{Vican:2012}
{Vican}, L. 2012, \aj, 143, 135

\bibitem[{{Wyatt} {et~al.}(2005){Wyatt}, {Greaves}, {Dent}, \&
  {Coulson}}]{Wyatt:2005}
{Wyatt}, M.~C., {Greaves}, J.~S., {Dent}, W.~R.~F., \& {Coulson}, I.~M. 2005,
  \apj, 620, 492

\bibitem[{{Wyatt} {et~al.}(2007){Wyatt}, {Smith}, {Greaves}, {Beichman},
  {Bryden}, \& {Lisse}}]{Wyatt:2007}
{Wyatt}, M.~C., {Smith}, R., {Greaves}, J.~S., {et~al.} 2007, \apj, 658, 569

\end{thebibliography}

\appendix
\section{Correcting the background bias between stars of different magnitudes}\label{AppA}

The best 5\% approach used in this paper to convert a series of individual null measurements to a single value creates a bias between stars of different magnitudes. The origin of this bias is the background noise which is presumably constant in absolute flux and divided by the flux of the star to compute the null. Therefore, the width of the relative background flux distribution decreases with the brightness of the star and the mean of the best 5\% measurements increases. The contribution of the background noise to the measured null then creates a bias that depends on the brightness of the star. In order to compensate for this effect, we can add a small fraction of background flux to the measured flux at null to match the relative background noise on stars of different magnitudes. Explicitly, we are looking for the fraction $x$ verifying the following equation:
\begin{equation}
{\rm variance} \left(\frac{B_2(t)}{F_2}\right) = {\rm variance} \left(\frac{B_1(t)}{F_1} + xB(t)\right)
\end{equation}
where $F_1$ and $F_2$ are the stellar fluxes measured in the photometric OBs (assuming $F_1>F_2$), $B_1(t)$ and $B_2(t)$ are the instantaneous background fluxes in the photometric aperture in the null OBs, and $B(t)$ is the instantaneous background flux measured in a nearby empty region of the detector. Assuming that $B_1(t)$ and $B(t)$ are uncorrelated, their variance adds up and the above equation becomes:
\begin{equation}
\frac{1}{F_2^2} = \frac{1}{F_1^2} +x^2 {\,\,\,\,\Rightarrow\,\,\,\,} x = \frac{\sqrt{F_1^2 - F_2^2}}{F_1F_2},\,\\
\end{equation}
where we have assumed that $\sigma_{B_1}$= $\sigma_{B_2}$= $\sigma_{B}$.

\end{document}